\documentclass[aps,prb,superscriptaddress,twocolumn,final,floatfix,showpacs]{revtex4-1}

\pdfoutput=1

\usepackage{amsmath}
\usepackage{graphicx}
\usepackage{latexsym}
\usepackage{units}
\usepackage{epstopdf}

\begin{document}

\title{ Lattice screening of the polar catastrophe and hidden 
        in-plane polarization in KNbO$_{3}$/BaTiO$_{3}$ interfaces}

\author{ Pablo Garc\'{\i}a-Fern\'andez }
\affiliation{ Departamento de Ciencias de la Tierra y
              F\'{\i}sica de la Materia Condensada, Universidad de Cantabria,
              Cantabria Campus Internacional,
              Avenida de los Castros s/n, 39005 Santander, Spain}
\author{ Pablo Aguado-Puente }
\affiliation{ Departamento de Ciencias de la Tierra y
              F\'{\i}sica de la Materia Condensada, Universidad de Cantabria,
              Cantabria Campus Internacional,
              Avenida de los Castros s/n, 39005 Santander, Spain}
\author{ Javier Junquera }
\affiliation{ Departamento de Ciencias de la Tierra y
              F\'{\i}sica de la Materia Condensada, Universidad de Cantabria,
              Cantabria Campus Internacional,
              Avenida de los Castros s/n, 39005 Santander, Spain}
\date{\today}

\begin{abstract}
 We have carried out first-principles simulations, based on 
 density functional theory, to obtain the atomic and electronic structure
 of (001) BaTiO$_{3}$/KNbO$_{3}$ interfaces in an isolated slab geometry.
 We tried different types of structures including symmetric and asymmetric
 configurations, 
 and variations in the thickness of the constituent materials.
 The spontaneous polarization of the layer-by-layer non neutral
 material (KNbO$_{3}$) in these
 interfaces cancels out almost exactly the ``built-in'' polarization
 responsible for the electronic reconstruction. 
 As a consequence, the so-called polar catastrophe is quenched and all 
 the simulated interfaces are insulating.
 A model, based on the modern theory of polarization and basic electrostatics,
 allows an estimation of the critical thickness
 for the formation of the two-dimensional electron gas between 
 42 and 44 KNbO$_{3}$ unit cells.
 We also demonstrate the presence of an unexpected in-plane polarization 
 in BaTiO$_{3}$
 localized at the $p$-type TiO$_2$/KO interface, 
 even under in-plane compressive strains.
 We expect this in-plane polarization to remain hidden due to angular 
 averaging during quantum fluctuations unless the symmetry is broken 
 with small electric fields.
\end{abstract}


\maketitle

\section{Introduction}
\label{sec:intro}

 The surprising discovery by Ohtomo and Hwang~\cite{Ohtomo-04} of a 
 metallic state at the interface between two good band insulating oxides, 
 LaAlO$_3$ and SrTiO$_3$, has triggered a large 
 amount of new studies on polar oxide interfaces.~\cite{Mannhart-08}
 Indeed, the quasi two-dimensional electron gas (2DEG) that forms
 when LaAlO$_{3}$ is grown on top of a TiO$_{2}$ terminated 
 (001)-surface of SrTiO$_{3}$, 
 i.e. when the interface between the two materials is LaO/TiO$_{2}$,
 displays very different properties from those generated at interfaces
 between standard III-V semiconductors (such as GaAs and 
 Al$_{x}$Ga$_{1-x}$As). 
 Among them we find 
 conducting carrier densities and electron effective masses
 orders of magnitude larger than those found at semiconductor 
 interfaces.~\cite{Mannhart-10}
 It is also fascinating how, depending on growth conditions,
 magnetic~\cite{Brinkman-07} and 
 superconducting~\cite{Reyren-07} ground states have been experimentally 
 identified at this interface between non-magnetic insulating oxides.
 Very recently, two independent groups have proven how both 
 magnetic and superconducting states might even coexist on the 
 same sample,~\cite{Li-11,Gardner-11} a very unexpected result since
 magnetic order is usually considered detrimental to superconductivity.
 As a consequence of all these phenomena, interfaces in polar oxides
 can open the door to novel implementations of field effect transistors
 and to a new era of oxide electronics.~\cite{Cen-09,Mannhart-10} 

 Despite this recent activity, many fundamental questions
 regarding the origin and confinement of the 2DEG remain highly debated.
 Different models have been proposed
 to explain the experimental results. 
 The pioneering one invokes the so called 
 ``polar catastrophe''~\cite{Nakagawa-06} 
 that arises from the polarization
 discontinuity~\cite{Stengel-09.4} between the III-III polar 
 LaAlO$_{3}$ film and the II-IV nonpolar SrTiO$_{3}$ layers along the [001]
 direction. 
 Indeed, from the formal ionic charge point of view, 
 LaAlO$_{3}$ can be described as a succession of positive 
 $\left({\rm La}^{+3} \rm{O}^{-2}\right)^{+1}$ 
 and negative $\left({\rm Al}^{+3} \rm{O}^{-2}_{2}\right)^{-1}$ layers,
 while the alternating $\left({\rm Sr}^{+2} \rm{O}^{-2}\right)^{0}$
 and $\left({\rm Ti}^{+4} \rm{O}^{-2}_{2}\right)^{0}$ layers
 of the perovskite structure of SrTiO$_{3}$ are charge neutral.
 But, aside this first rationalization, other explanations can be
 found in the literature for the origin of the 2DEG, among them:
 (i) the interlayer mixing between LaAlO$_{3}$ and SrTiO$_{3}$ 
 and non-abruptness of the interfaces
 (with the formation of a few monolayers of metallic 
 La$_{1-x}$Sr$_{x}$TiO$_{3}$~\cite{Willmott-07}), 
 (ii) doping due to oxygen vacancies~\cite{Herranz-07,Kalabukhov-07} 
 (including those produced in surface redox reactions~\cite{Bristowe-11}),
 and/or (iii) the presence of charged defects and
 adsorbates.~\cite{Cen-08,Son-10}
 All these models highlight the importance of the growth conditions of
 these structures for the appearance and the behavior
 of the functional properties of the 2DEG.

 One of these properties, that is well reproduced by different experimental
 groups on many samples grown with a variety of techniques, 
 is the existence of a critical thickness, $t_{\rm{c}}$, in the number
 of layers of LaAlO$_{3}$ for the formation of the 2DEG.
 Thiel and coworkers~\cite{Thiel-06} have demonstrated that, 
 for the interfaces to be conducting, the number of layers of LaAlO$_{3}$ 
 has to be larger than four unit cells. 
 The thickness of the polar layer increases up to five unit cells in $n$-type 
 LaVO$_{3}$/SrTiO$_{3}$ interfaces.~\cite{Hotta-07}
 These observations are consistent with the fact that the conductivity of 
 SrTiO$_{3}$-LaAlO$_{3}$-SrTiO$_{3}$ heterostructures 
 with dissimilar interfaces is reduced if 
 their $p$-type (AlO$_{2}$/SrO) and 
 $n$-type (LaO/TiO$_{2}$) interfaces are spaced by less than six unit
 cells.~\cite{Huijben-06,Chen-09}
 Remarkably, even below the critical thickness, a metal-insulator transition
 can be driven by an external electric field.~\cite{Thiel-06,Cen-08}
 These studies suggested the possibility of the design of new polar interfaces 
 where the appearance of the 2DEG could be switched on and off by the
 action of an external perturbation.

 Along this line, the replacement of one (or the two) materials
 at the interface by ferroelectric perovskites is particularly attractive. 
 The spontaneous polarization present in these 
 materials is very sensitive to electric fields and could 
 be used to create a bound charge 
 at the interface that could reinforce/deplete the 2DEG.
 Previous theoretical works have been focused 
 on I-V/II-IV interfaces [NaNbO$_{3}$/SrTiO$_{3}$,~\cite{Oja-09,Oja-12}
 and KNbO$_{3}$/$A$TiO$_{3}$
 interfaces, where $A$ = Sr, Ba or Pb~\cite{Niranjan-09,YWang-09,Oja-12}].
 From the formal ionic charge point of view, 
 I-V ferroelectric perovskite oxides, such as NaNbO$_{3}$ and KNbO$_{3}$,
 are made of alternating 
 positive (B$^{+5}$O$^{-2}_2$)$^{+1}$ and negative (A$^{+1}$O$^{-2}$)$^{-1}$ 
 charged layers along the [001] direction
 (essentially as LaAlO$_{3}$, although now the AO layers of the 
 perovskite ABO$_{3}$ structure  
 are negative, while the BO$_{2}$ layers are positive).
 Therefore, the layer-by-layer electrostatic of the
 previous I-V/II-IV interfaces
 is analogous to that in the LaAlO$_{3}$/SrTiO$_{3}$ interface.
 Density functional theory simulations on non-stoichiometric 
 (i.e. with a non-integer number of unit cells 
 of the layer-by-layer non neutral perovskite), symmetric
 superlattices indeed suggested the existence
 of a 2DEG in KNbO$_{3}$/$A$TiO$_{3}$ interfaces, 
 switchable between two conducting states by the 
 ferroelectric polarization orientation of the titanate 
 layer.~\cite{Niranjan-09,YWang-09}

 However, with the simulation boxes used in the previous works only
 the $n$-type or $p$-type interfaces are present.
 It can be proved (see Sec. 4 of the Supplemental Material of 
 Ref.~\onlinecite{Stengel-11.3})
 that, within this configuration, the local interface
 properties exactly reproduce those of the infinite isolated slab 
 geometries, obviously beyond the critical thickness 
 for the formation of the 2DEG.
 In other words, the calculations of Refs.~\onlinecite{Niranjan-09} and 
 Ref.~\onlinecite{YWang-09} show
 the charge distribution and properties after the
 electronic reconstruction has taken place,
 but nothing is said about the magnitude of the critical thickness
 for the formation of the 2DEG.~\cite{Lee-08}
 
 In this work we carry out first principles calculations of 
 BaTiO$_{3}$/KNbO$_{3}$ interfaces where we explicitly avoid  
 the issue of lack of stoichiometry in the simulation box.
 We have found that, due to the large lattice screening provided by
 the KNbO$_{3}$ layer, the critical thickness for the formation  
 of the 2DEG is one order of magnitude larger than in LaAlO$_{3}$/SrTiO$_{3}$
 interfaces.
 An unexpected result of our simulations is that an in-plane
 polarization develops on the BaTiO$_3$ side of a 
 (TiO$_{2}$/KO) $p$-type BaTiO$_3$/KNbO$_3$ interface even 
 when BaTiO$_{3}$ is subject to in-plane compressive strains.
 We explain this effect using basic electrostatic arguments.

 The rest of the paper is organized as follows. 
 After summarizing the basic theory behind the electronic reconstruction
 in Sec.~\ref{sec:summaryer},
 we present the computational details used in our simulations 
 in Sec.~\ref{sec:computationaldetail}.
 The first-principles results, together with the relevant comparisons
 to the model, can be found in Sec.~\ref{sec:results}.
 
\section{Background on the ``polar discontinuity" model}
\label{sec:summaryer}

 In order to establish the nomenclature and the basic theory that 
 will be used later,
 we review the most important points of the ``polar discontinuity'' model.
 Although this model has been invoked since the discovery of the 2DEG 
 at polar oxide interfaces,~\cite{Nakagawa-06} 
 only recently it has been rigorously rationalized 
 with explanations firmly rooted on the modern theory of 
 polarization (for a recent review, see Ref.~\onlinecite{Resta-07} and 
 references therein).
 This has been developed by Stengel and Vanderbilt in 
 Ref.~\onlinecite{Stengel-09.4} for insulating interfaces,
 and later generalized by Stengel for the case of a non-zero surface density
 of ``free'' charge in Ref.~\onlinecite{Stengel-11.2}, 
 and to the case of surfaces in Ref.~\onlinecite{Stengel-11.3}.
 The theory presented in these works is absolutely general, and
 we strongly point the interested reader to those
 milestone papers.
 Here, we particularize it to the conditions considered in this work, 
 and estimate the critical thickness for the formation of 2DEG in the
 case where any of the two materials forming the interface is ferroelectric.

 The standard nomenclature used in the literature of polar oxide
 interfaces denote the material that is non-neutral layer-by-layer
 (i.e. LaAlO$_{3}$) as \emph{polar},
 and the material that is neutral layer-by-layer (i.e. SrTiO$_{3}$)
 as \emph{non-polar}.
 Rigorously, this notation does not apply here since the
 two materials that constitute our interfaces are ferroelectric
 and, therefore, might undergo polar phase transitions 
 with the appearance of a non-vanishing spontaneous polarization.
 Nevertheless, for the sake of consistency with previous works
 we will maintain the convention and refer to the
 ferroelectric non-neutral layer by layer material (i.e. KNbO$_{3}$) as 
 the polar material and  
 the ferroelectric neutral layer-by-layer material (i.e. BaTiO$_{3}$)
 as the non-polar one.

\begin{figure} [h]
    \begin{center}
       \includegraphics[width=8.5cm]{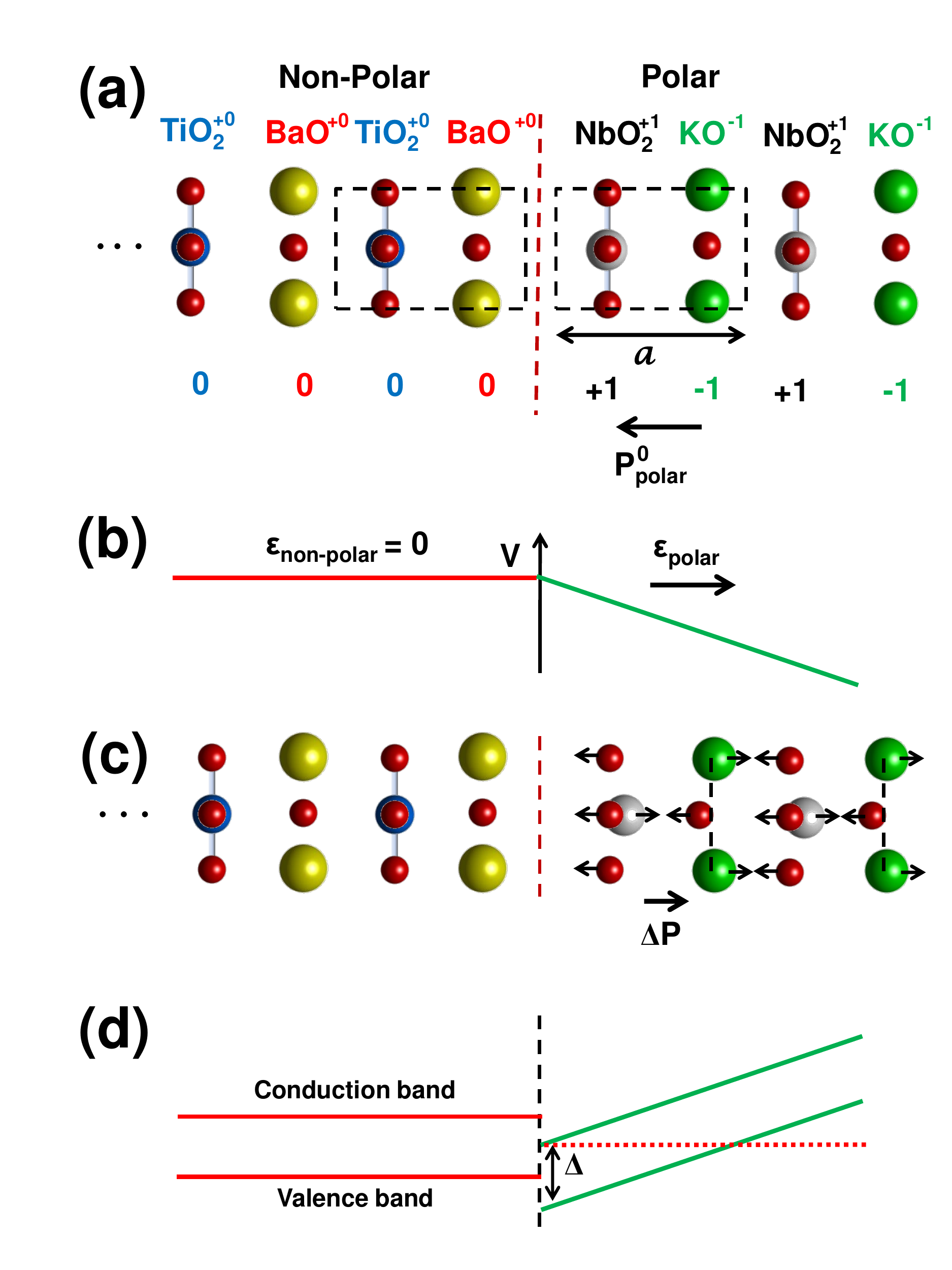}
       \caption{ (Color online) 
                 (a) Schematic representation of a pristine polar interface
                 in an isolated slab geometry. Atoms are represented by balls:
                 O (small size, red), Ba (large size, yellow), 
                 Ti (medium size, blue), K (large size, green), and
                 Nb (medium size, gray), with the corresponding atomic layer
                 printed on top. 
                 The dots at the left of the BaTiO$_{3}$ indicates that we
                 assume a thick layer. 
                 At the other side, a free KO-terminated surface 
                 of the polar KNbO$_{3}$ is assumed.
                 Numbers below each layer indicate the formal ionic charge.
                 The interface is marked with a red dashed line. 
                 Our choice for the unit cells that tile the entire crystal 
                 are represented by black dashed boxes. 
                 $a$ stands for the out-of-plane lattice constant of the 
                 polar material unit cell.
                 (b) Electrostatic potential of the polar interface in the
                 isolated slab geometry. Within our electrostatic boundary
                 conditions, the macroscopic electric field in BaTiO$_{3}$
                 is forced to be zero. Note that the field within the 
                 non-neutral layers points away from the interface.
                 (c) Induced polarizations $\Delta P$ due to the macroscopic
                 electric fields.
                 (d) Schematic representation of the energy bands of the polar
                 interface.}
       \label{fig:scheme}
   \end{center}
 \end{figure}

 During the development of the model we assume a $n$-type interface,
 simulated within an isolated slab geometry,
 with the non-polar material at the left
 and the polar material (with a formal ionic charge of 
 $\pm e$ alternating from layer to layer, where
 $e$ is the magnitude of the electronic charge)
 at the right [see Fig.~\ref{fig:scheme}(a)]. 
 The generalization for other configurations is straightforward, 
 changing the appropriate signs when required.
 
 Within the modern theory of polarization, we can compute the
 ``formal'' bulk polarization from the positions of the atomic nuclei
 and the center of localized Wannier functions.
 This decomposition of the charge (nuclear and electronic) into localized
 contributions allows for a simple classical interpretation of the bulk
 polarization in terms of a point charge model, and rescue the
 Clausius-Mossotti formulation.

 In the perfectly ideal structure without rumpling, 
 where an atomically sharp junction in the absence of defects is supposed,
 all the atoms at a given layer lie at the same plane 
 [Fig.~\ref{fig:scheme}(a)].
 Then, we can always choose unit cells that tile the crystal
 under appropriate primitive translations, and that leave the left-over
 interface region charge neutral.~\cite{Stengel-09.4}
 (It is important to note that, for the moment,
 we are assuming that the thickness of
 the polar layer is below
 the critical thickness for the formation of the 2DEG.)
 Then, the magnitude of the dipole of an individual (AO)-(BO$_{2}$) unit  
 in this material is $d = e a/2$, where $a$ is the out-of-plane 
 lattice constant.
 The sign of the dipole is always directed from the negative to the positive 
 layer, so in the considered configuration the sign is negative 
 [see Fig.~\ref{fig:scheme}(a)].
 This dipole corresponds to a ``built-in formal'' 
 polarization (calculated for the 
 previous choice of unit cell for the primitive basis of atoms and Wannier
 functions) of

 \begin{equation}
    P^{0}_{\rm polar} = 
     \frac{d}{\Omega} = - \frac{ea}{2aS} = - \frac{e}{2S},
    \label{eq:formalpol}
 \end{equation}

 \noindent where $\Omega$ is the volume of the unit cell of the
 polar material and $S$ is the cell surface. 
 Analogously, since the layers are formally charged neutral in the non-polar
 material, $P^{0}_{\rm non-polar} = 0$.

 Now, we can wonder what would happen if the materials
 that constitute the interface are ferroelectric, with a
 non-vanishing ferroelectric contribution to the polarization $P^{\rm S}$
 (this might be also the case when a compressive epitaxial strain is applied
 to the SrTiO$_{3}$/LaAlO$_{3}$ interface~\cite{Bark-11}).
 In this situation, the ferroelectric polarization
 must be added to the ``built-in'' polarization of
 the polar layer.~\cite{Stengel-11.3}

 The difference in the polarization at the interface between the
 two materials produces a surface density of bound charge, 

 \begin{equation}
    \sigma_{\rm bound} = P_{\rm non-polar} - P_{\rm polar}.
    \label{eq:sigma}
 \end{equation}

 \noindent Classical electromagnetic theory (Gauss's theorem) 
 teaches us that this 
 sheet of interfacial charge gives rise to a change
 in the macroscopic electric field in the two materials of 

 \begin{equation}
    \mathcal{E}_{\rm polar} - \mathcal{E}_{\rm non-polar} = 
    \frac{\sigma_{\rm bound}}{\epsilon_{0}},
    \label{eq:efield}
 \end{equation}

 \noindent where $\epsilon_{0}$ is the dielectric permittivity of vacuum
 [Fig.~\ref{fig:scheme}(b)].
 The exact magnitude of the fields depends on the electrostatic 
 boundary conditions, extremely linked with the geometry of the simulation
 boxes used in the computations 
 (see Supplemental Material of Ref.~\onlinecite{Stengel-11.2} 
 for a complete review).

 The field might induce strong structural changes in the materials and polarizes
 the electronic Wannier functions. Both facts translate into 
 the development of a 
 field-induced polarization $\Delta P$ 
 [Fig.~\ref{fig:scheme}(c)] that tends to screen the
 discontinuity of the total polarization $P$,

 \begin{equation}
    P = P^{0} + P^{\rm S} + \Delta P
    \label{eq:totalpol}
 \end{equation}

 \noindent and, consequently, of the macroscopic electric field.
 At the end, a self-consistent solution of 
 Eqs.~(\ref{eq:sigma})-(\ref{eq:totalpol}) is achieved 
 where the polarizations
 in the two materials are in equilibrium with the corresponding 
 macroscopic electric fields, and the total energy of the
 system is minimized.

 For a sufficiently thin polar layer thickness, before the electronic
 reconstruction takes place, the absence of free charge at the interface
 requires the normal component of the electric
 displacement field $D$ to be preserved,

 \begin{equation}
    D_{\rm non-polar} = D_{\rm polar} \equiv D.
    \label{eq:continuityd}
 \end{equation}

 \noindent The finite electric displacement is an input parameter of the model.
 In a first-principles simulations its value can be set by hand
 using the virtual crystal approximation to introduce external
 fractional charges in the surface atoms layers, while constraining
 the macroscopic electric field to be strictly zero in the vacuum 
 region.~\cite{Stengel-11.2}
 Other authors fix the atomic positions on the surface unit cell
 to some specific values.~\cite{Bark-11}

 From the definition of the electric displacement field,

 \begin{align}
    D_{\rm polar} & = \epsilon_{0} \mathcal{E}_{\rm polar} + 
                            P_{\rm polar}
    \nonumber \\
                  & = \epsilon_{0} \mathcal{E}_{\rm polar} + 
                      P^{0}_{\rm polar} + P^{\rm S}_{\rm polar} + 
                      \Delta P_{\rm polar} ,
    \label{eq:contd}
 \end{align}

 \noindent and assuming that the polar material behaves as a linear dielectric
 of susceptibility $\chi_{\rm polar}$ around the spontaneous polarization
 structure, then

 \begin{equation}
    \Delta P_{\rm polar} = \epsilon_{0} \chi_{\rm polar} 
                           \mathcal{E}_{\rm polar}, 
    \label{eq:ppolar}
 \end{equation}

 \noindent and Eq.~(\ref{eq:contd}) transforms into

 \begin{equation}
    D_{\rm polar}   = \epsilon_{0} \epsilon_{\rm polar}
                      \mathcal{E}_{\rm polar} + 
                      P^{\rm S}_{\rm polar} + P^{0}_{\rm polar},   
    \label{eq:contd2}
 \end{equation}

\noindent where $\epsilon_{\rm polar} = 1 + \chi_{\rm polar}$ 
 is the dielectric
 constant of the polar material.  
 From Eqs.~(\ref{eq:continuityd}) and ~(\ref{eq:contd2}),

 \begin{equation}
    \mathcal{E}_{\rm polar} = \frac{D -
                      \left( P^{0}_{\rm polar} + 
                             P^{\rm S}_{\rm polar} \right)}
                             {\epsilon_{0} \epsilon_{\rm polar}} .
    \label{eq:fieldpol}
 \end{equation}

 This electric field tilts the electronic bands of the polar layer
 [Fig.~\ref{fig:scheme}(d)]. At a given 
 critical thickness, $t_{\rm{c}}$, the top of the valence band 
 of the polar material reaches the level of the bottom of the
 conduction bands.~\cite{Reinle-Schmitt-11}
 Beyond $t_{\rm c}$, a Zener breakdown takes place,
 with the concomitant transfer of 
 charge from the surface of the polar
 material to the interface.  
 The magnitude of $t_{\rm c}$ can be easily computed from 
 Eq.~(\ref{eq:fieldpol}) as

 \begin{equation}
    t_{\rm c} = \frac{\Delta}{ e \left| \mathcal{E}_{\rm polar} \right| }
              = \frac{\epsilon_{0} \epsilon_{\rm polar} \Delta}
                     { e \left| D - 
                      \left(P^{0}_{\rm polar} + P^{\rm S}_{\rm polar} 
                      \right) \right|} , 
    \label{eq:crithick}
 \end{equation}
 
 \noindent where $\Delta$ is the interfacial potential step.
 $\Delta$ will depend on the type of band alignment and
 on the particular interface ($p$ or $n$).
 Here, according to the band alignment of Fig.~\ref{fig:scheme}(d) 
 for the $n$ interface is given by~\cite{Deltap}
   
 \begin{equation}
    \Delta =E_{\rm gap}^{\rm polar}, 
    \label{eq:delta}
 \end{equation}

 \noindent where $E_{\rm gap}^{\rm polar}$ is the band-gap of the
 polar material. 
 Note that the band alignment in our interfaces, where
 the band gaps of BaTiO$_{3}$ and KNbO$_{3}$ are very similar, 
 is of type II, 
 different from that in the prototypical LaAlO$_{3}$/SrTiO$_{3}$ case
 (type I). For schematic views of the types of interfaces   
 according to the band offset see Ref.~\onlinecite{wiki-heterojunction}.

 The first conclusion that can be drawn from Eq.~(\ref{eq:crithick}) 
 is that the polarizability of the polar layer
 is essential to determine the 
 critical thickness for the formation of the 2DEG
 (it is directly proportional to $\epsilon_{\rm polar}$).
 The role played by the polar distortions to avoid the polar catastrophe
 in LaAlO$_{3}$/SrTiO$_{3}$ interfaces has been confirmed by first-principles
 simulations.~\cite{Pentcheva-09}
 The importance of the extra screening due to lattice relaxations
 has also been discussed in the strongly related
 LaTiO$_{3}$/SrTiO$_{3}$ interfaces.~\cite{Hamann-06,Okamoto-06,Larson-08}
 Indeed, the success for explaining the critical thickness for the formation
 of the 2DEG,~\cite{Thiel-06,Lee-08,Chen-09} 
 together with the electrostrictive effect
 on the polar LaAlO$_{3}$ films,~\cite{Cancellieri-11}
 are between the most important achievements of the polar discontinuity model.

 The second conclusion is related with the relationship between 
 $D$ and the ferroelectric polarization in the non-polar layer.
 In many works, the simulations try to reproduce the behavior
 of a thin polar layer on top of a thick non-polar substrate.
 In such cases, the macroscopic field in the non-polar materials is forced
 to be zero, either by symmetry or by using a dipole correction in
 vacuum. 
 As a consequence, $D = P_{\rm non-polar}$.
 If the ferroelectric contribution to the polarization in the non-polar
 layer points in the same direction as the ``built-in'' polarization
 in the polar one, then it contributes to increase the 
 critical thickness.
 This has been proven in 
 Ref.~\onlinecite{Bark-11} for the case of a
 SrTiO$_{3}$/LaAlO$_{3}$ interface subject
 to epitaxial strain [Eq.~(\ref{eq:crithick}) is equivalent to Eq. (4) in 
 Ref.~\onlinecite{Bark-11} taking into account that, in this particular
 system $P^{\rm S}_{\rm polar} = P^{\rm S}_{\rm LaAlO_{3}} = 0$
 and $D = P^{\rm S}_{\rm SrTiO_{3}}$].

 Finally, the third conclusion is that there is also a strong influence
 of an eventual 
 spontaneous polarization of the polar material in the value of $t_{\rm c}$.
 In particular, if $P^{\rm{S}}_{\rm polar}$ is close in magnitude 
 to $P^{0}_{\rm polar}$ and points in the opposite direction, 
 so the ``built-in'' polarization can be almost compensated by the
 ferroelectric contribution to the polarization, 
 a large cancellation of the term
 in parentheses in the denominator of Eq.~(\ref{eq:crithick}) is produced,
 with the concomitant increase in the critical thickness.
 In a previous work, Murray and Vanderbilt~\cite{Murray-09} have estimated
 how in SrTiO$_{3}$/KNbO$_{3}$ \emph{superlattices} the system would
 not become metallic until the number of layers of KNbO$_{3}$ is larger
 than 32.

 To further validate Eq.~(\ref{eq:crithick}) 
 in an \emph{isolated slab} geometry,
 we have carried out simulations on BaTiO$_{3}$/KNbO$_{3}$ interfaces.
 The motivation for this choice is four fold:
 (i) under appropriate compressive in-plane strains, 
 KNbO$_{3}$ is a ferroelectric polar material, with 
 the spontaneous polarization pointing along the [001] direction,
 and with a ``built-in'' polarization of 
 $P^{0}_{\rm polar} =P^{0}_{\rm KNbO_{3}} = 
 e/2S \approx 53 \:\: \mu {\rm C/cm}^{2}$ 
 (value computed at the theoretical in-plane 
 lattice constant of an hypothetical SrTiO$_{3}$ substrate,
 $a_{\parallel}$ = 3.874 {\AA}),
 (ii) under this mechanical boundary condition, 
 both KNbO$_{3}$ and BaTiO$_{3}$ can be stabilized with the same tetragonal
 $P4mm$ symmetry, 
 (iii) the theoretical spontaneous polarization of our tetragonal KNbO$_{3}$ 
 in bulk is 
 $P^{\rm S}_{\rm polar} = P^{\rm S}_{\rm KNbO_{3}} = 
 48 \:\: \mu {\rm C/cm}^{2}$,
 close to the built-in polarization, and
 (iv) we can also test to which extent the ferroelectric contribution to the
 polarization in the BaTiO$_{3}$ layer is dominated by the imposed
 value of the displacement field in the simulations.
 First-principles results and comparison with the previous model
 will be presented in Sec.~\ref{sec:results}.

\section{Computational details}
\label{sec:computationaldetail}

 We have carried out density functional first-principles simulations based
 on a numerical atomic orbital method as implemented in the {\sc Siesta}
 code.~\cite{Soler-02}
 All the calculations have been carried out within the
 local density approximation (LDA), using the Perdew and Zunger~\cite{Perdew-81}
 parametrization of the Ceperley and Alder functional~\cite{Ceperley-80}
 to simulate the electronic exchange and correlation.
 This choice avoids the systematic overestimation of the 
 ferroelectric character~\cite{Junquera-08}
 of perovskite oxides found in other commonly used 
 functionals~\cite{Perdew-96} based
 on the generalized gradient approximation. 
 This is an important point in this study, 
 since the dielectric properties of the oxides at the bulk level
 might determine the behavior of the interfaces, in particular
 a tendency for ``overscreening'' of the 2DEG
 when the ferroelectric properties are favored.~\cite{Stengel-11.2}

 Core electrons were replaced by {\it ab-initio} norm conserving
 pseudopotentials, generated using the
 Troullier-Martins scheme,~\cite{Troullier-91} in the
 Kleinman-Bylander fully non-local separable representation.~\cite{Kleinman-82}
 Due to the large overlap between the semicore and valence states,
 the semicore $3s$ and $3p$ electrons of Ti,
 $3s$ and $3p$ electrons of K,
 $4s$ and $4p$ electrons of Nb, and
 $5s$ and $5p$ electrons of Ba
 were considered as valence electrons
 and explicitly included in the simulations.
 K, Ti, Nb and Ba pseudopotentials were generated scalar relativistically.
 The reference configuration and cutoff radii for each angular momentum shell
 for the pseudopotentials used
 in this work
 can be found in Ref.~\onlinecite{Junquera-03.2} for Ba, Ti, and O,
 and in Table~\ref{table:pseudopotentials} for K and Nb.

 \begin{table*}
    \caption[ ]{ Reference configuration and cutoff radii
                 of the pseudopotential used in our study. Because
                 of the inclusion of the semicore states in valence within
                 the Troullier-Martin scheme, K and Nb pseudopotentials
                 must be generated for ionic configurations (ionic
                 charge of +1). However, these are
                 more suitable than the neutral ones, given the
                 oxidation numbers of these atoms in the perovskites.
                 Units in Bohr.
               }
    \begin{center}
       \begin{tabular}{ccccc}
          \hline
          \hline
                        &
                        &
          K             &
          Nb            \\
          Reference                         &
                                            &
          $3s^{2}, 3p^{6}, 3d^{0}, 4f^{0}$  &
          $4s^{2}, 4p^{6}, 4d^{4}, 4f^{0}$  \\
          \hline
          Core radius                       &
          $s$                               &
          1.50                              &
          1.45                              \\
                                            &
          $p$                               &
          1.35                              &
          1.50                              \\
                                            &
          $d$                               &
          1.50                              &
          1.40                              \\
                                            &
          $f$                               &
          2.00                              &
          2.00                              \\
          Scalar relativistic?              &
                                            &
          yes                               &
          yes                               \\
          \hline
          \hline
       \end{tabular}
    \end{center}
    \label{table:pseudopotentials}
 \end{table*}

 The one-electron Kohn-Sham eigenstates were expanded in a basis of
 strictly localized numerical atomic orbitals.~\cite{Sankey-89,Artacho-99}
 We used a single-$\zeta$ basis set for the semicore states of 
 K, Ti, Nb, and Ba and double-$\zeta$ plus polarization for the valence
 states of all the atoms. For K (Ba), an extra shell of 3$d$ (5$d$) 
 orbitals was added.
 All the parameters that define the shape and range of the basis functions
 were obtained by a variational
 optimization of the energy~\cite{Junquera-01} in bulk cubic BaTiO$_{3}$ 
 (for Ba, Ti, and O),
 and of the entalphy~\cite{Anglada-02} (with a pressure P = 0.03 GPa) 
 in bulk cubic KNbO$_{3}$ 
 (for K and Nb, the basis set of O was frozen to that obtained in BaTiO$_{3}$). 

 The electronic density, Hartree, and exchange correlation potentials,
 as well as the corresponding matrix elements between the basis orbitals,
 were calculated in a uniform real space grid. An equivalent plane
 wave cutoff of 1200 Ry was used to represent the charge density.
 For the Brillouin zone integrations we use a
 Monkhorst-Pack sampling~\cite{Monkhorst-76} equivalent to
 $12 \times 12 \times 12$ in a five atom perovskite unit cell.

 To avoid the problem of artificially charging the interface with 
 the use of non-stoichiometric superlattices with periodic boundary conditions,
 we follow the proposal of Lee and Demkov.~\cite{Lee-08} 
 Within this approach, the calculations were performed
 on vacuum-terminated (KNbO$_{3}$)$_m$/(BaTiO$_{3}$)$_l$/(KNbO$_{3}$)$_m$ slabs,
 where the number of KNbO$_{3}$ cells, $m$, is always an integer number to 
 guarantee the  electrical neutrality of the interface. 
 In particular, we have focused on two kinds of systems: 
 symmetric slabs where both interfaces between BaTiO$_{3}$ and KNbO$_{3}$ 
 are of the same kind (either $p$-type TiO$_{2}$/KO, or 
 $n$-type BaO/NbO$_{2}$ interfaces), and asymmetric interfaces where one 
 interface is TiO$_2$/KO and the other BaO/NbO$_2$. 
 In both cases we have relaxed structures containing a different number of 
 KNbO$_{3}$ and BaTiO$_{3}$ unit cells, given by the subscripts $m$ and $l$, 
 respectively. 
 In the asymmetric interface a dipole slab correction is used
 to guarantee that the electric field in vacuum vanishes.

 The in-plane lattice constant was fixed to the theoretical one of an
 hypothetical SrTiO$_{3}$ substrate ($a_{\parallel}$ = 3.874 {\AA}).
 Under this constraint, both BaTiO$_{3}$ and KNbO$_{3}$ are under
 conditions of compressive epitaxial strains. This will be relevant 
 for the discussion of Sec.~\ref{sec:in-plane}.

 Starting from ideal reference structures,
 built by piling up the corresponding unit cells of bulk strained materials
 without rumpling, the atomic coordinates are relaxed 
 until the maximum component of the force on any atom was smaller than
 0.01 eV/{\AA}.
 In the symmetric slabs, the minimization is performed in a two step process:
 First, mirror symmetry planes are imposed on the central layer of 
 BaTiO$_{3}$ to avoid the development of a polarization in
 any direction. 
 Then, symmetry is broken displacing coherently 
 the cations by hand,
 and a second relaxation is carried out without any imposed symmetry.
 In the asymmetric slabs, the relaxations are carried out in a single step,
 since the symmetry is broken directly in the initial reference structure.
 
 To stablish the notation, we will call the plane parallel to the interface
 the $(x,y)$ plane, whereas the perpendicular direction will be referred to
 as the $z$-axis.

 \section{Results}
 \label{sec:results}

 \subsection{Out-of-plane lattice relaxations and screening} 
 \label{sec:oop-relaxation}

\begin{figure} [h]
    \begin{center}
        \includegraphics[width=8.5cm]{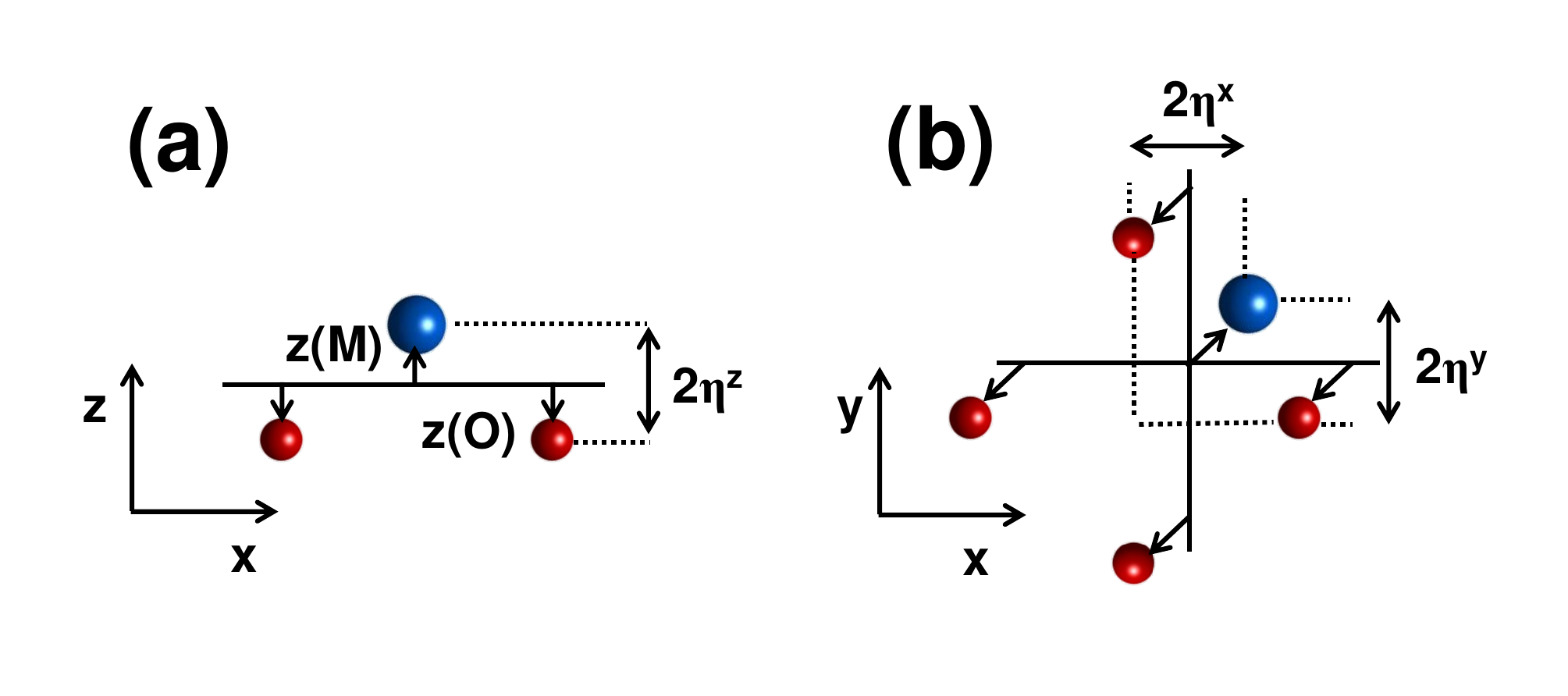}
        \caption{(Color online) Schematic representation about the
                 procedure to calculate 
                 (a) the out-of-plane rumpling, $\eta^{z}$, and
                 (b) the in-plane rumplings, $\eta^{x}, \eta^{y}$.
                 Solid lines represent the position of the atoms in the
                 ideal unrelaxed structure. After relaxation, the atoms
                 move in the directions indicated by the arrows.
                 Meaning of the balls as in Fig.~\ref{fig:scheme}. }
        \label{fig:illurump}
    \end{center}
 \end{figure}

 In order to characterize the atomic displacements induced by 
 the relaxation, we define the ``out-of-plane'' 
 rumpling parameter along $z$ of layer 
 $i$ as $\eta_{i}^{z} = \left[ z (M_{i}) - z (O_{i})\right]/2$,
 where $z (M_{i})$ and $z (O_{i})$ are, respectively, the 
 $z$ coordinates of the cations and the oxygens at a given layer $i$ 
 [Fig.~\ref{fig:illurump}(a)].
 We also define the ``in-plane'' rumplings ($\eta^{x}, \eta^{y}$),
 in an equivalent way, as represented graphically in 
 Fig.~\ref{fig:illurump}(b). 

\begin{figure} [h]
    \begin{center}
        \includegraphics[width=8.5cm]{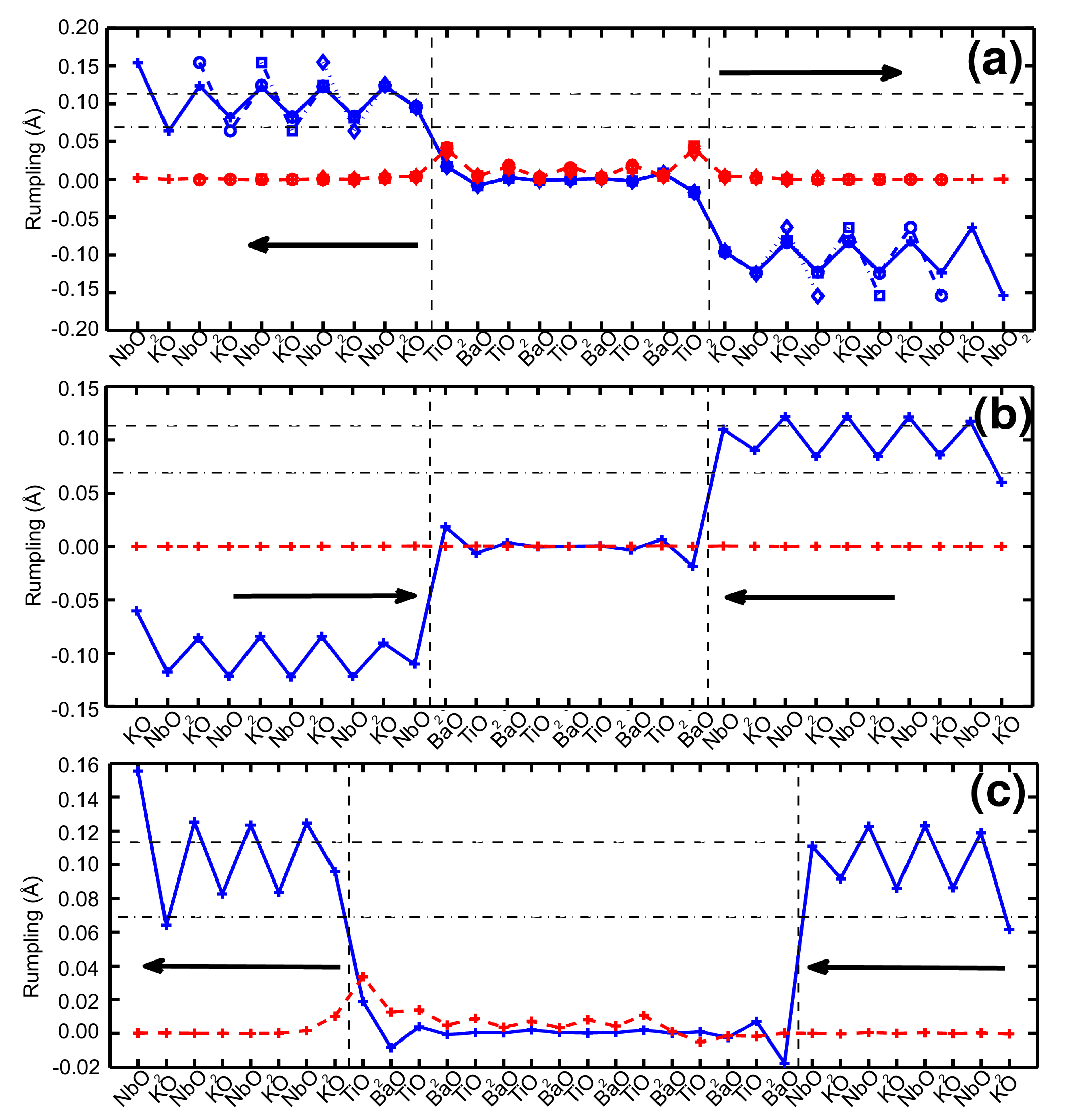}
        \caption{(Color online) Out-of-plane (solid blue) 
                 and in-plane (long-dashed red) 
                 lattice polarization calculated for 
                 (a) symmetric-$p$ [$l$ fixed to 4.5 and
                 $m$ = 2 (diamonds), 
                 $m$ = 3 (squares), 
                 $m$ = 4 (circles),
                 $m$ = 5 (crosses)], 
                 (b) symmetric-$n$ ($m$=5, $l$ = 4.5),
                 and (c) asymmetric ($m$=4, $l$=8)  
                 BaTiO$_{3}$/KNbO$_{3}$ slabs.
                 Short-dashed (dot-dashed) lines represent
                 the rumplings of the NbO$_{2}$ (KO) layers
                 in bulk KNbO$_{3}$ under the same epitaxial constraint.
                 The arrows point along the direction of the 
                 ``built-in'' polarization.
                 Vertical dashed lines indicate the position of the interfaces.
                 The in-plane rumplings plotted here correspond to
                 $\eta_{xy} = \sqrt{\eta_{x}^{2} + \eta_{y}^{2}}$ 
                 defined in Fig.~\ref{fig:illurump}.
                 }        
        \label{fig:rumpling}
    \end{center}
 \end{figure}

 In all the studied systems a large out-of-plane rumpling 
 is observed within the KNbO$_{3}$ layers,
 with a magnitude that is essentially independent of their thickness
 and the kind of interface: symmetric $p$ [Fig.~\ref{fig:rumpling}(a)], 
 symmetric $n$ [Fig.~\ref{fig:rumpling}(b)]
 or asymmetric [Fig.~\ref{fig:rumpling}(c)].
 Two oxide layers away from the interface, the layer-by-layer
 rumpling converges to a rather uniform sawtooth pattern, 
 with values slightly larger than those observed in bulk KNbO$_{3}$
 under the same epitaxial conditions.
 This fact is consistent with 
 $\left| P^{0}_{\rm KNbO_{3}}\right| > \left| P^{\rm S}_{\rm KNbO_{3}}\right|$,
 with $\Delta P_{\rm KNbO_{3}}$ tending to compensate 
 for the difference in order to screen the polarization
 discontinuity at the interface.
 Only in the neighborhood of the free surface, a small deviation 
 from this trend is obtained due to the larger relaxations
 on the surface atoms. 

 The dipole slab correction ensures that the displacement
 field in vacuum vanishes, $D = 0$. 
 Due to the absence of surface external charges in the simulations,
 this value is preserved at the interfaces.
 This implies that a polarization in the BaTiO$_{3}$ layer
 induces a depolarizing field that is responsible for a large
 electrostatic energy. Therefore, under this electrical
 boundary conditions, no polarization is expected
 on BaTiO$_{3}$.
 Both facts are well reproduced in our simulations,
 where we observe that the out-of-plane polarization vanishes within the 
 BaTiO$_{3}$ layer in all cases. This happens even when the mirror
 symmetries are lifted by hand (in the symmetric interfaces),
 or spontaneously (in the asymmetric slabs).

\begin{figure} [h]
    \begin{center}
       \includegraphics[width=10.0cm]{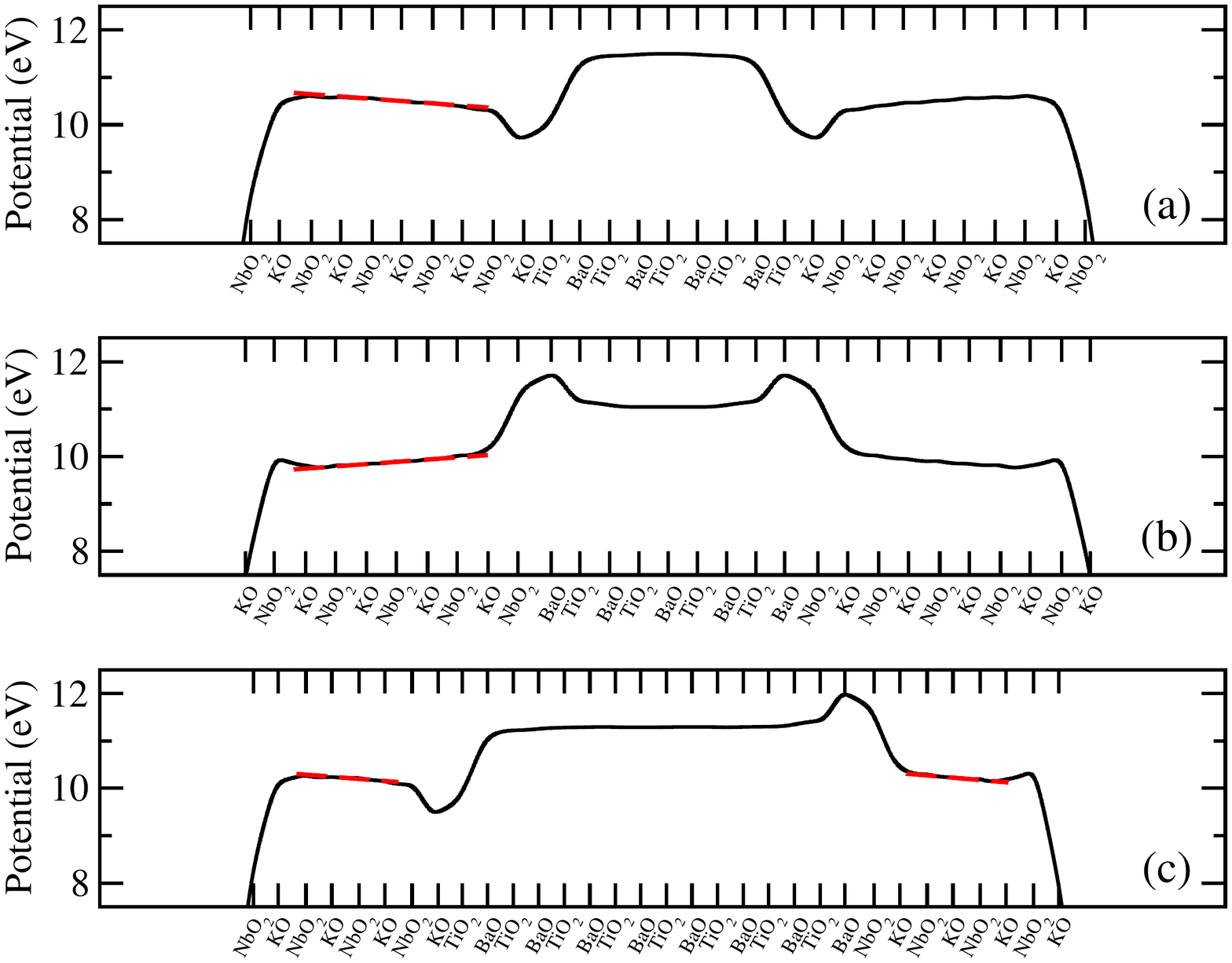}
       \caption{ (Color online) 
                 Macroscopically averaged internal electrostatic potential 
                 (black solid line)
                 for
                 (a) symmetric-$p$ ($m$ = 5, $l$ = 4.5), 
                 (b) symmetric-$n$ ($m$ = 5, $l$ = 4.5),
                 and (c) asymmetric ($m$=4, $l$=8)  
                 BaTiO$_{3}$/KNbO$_{3}$ slabs.
                 The red dashed lines indicates the region used to 
                 extract numerically the macroscopic field.
               }
        \label{fig:potential}
    \end{center}
 \end{figure}

 According to the model developed in Sec.~\ref{sec:summaryer}, 
 the polarization in the polar layer
 tends to screen the discontinuity of the
 polarization at the interface and, therefore, its sign opposes 
 that of the ``formal'' polarization.
 The lattice screening avoids the development
 of an electric field in the polar material that would result in the
 tilting of its bands and,
 for sufficiently large thicknesses, to a Zener breakdown and 
 accumulation of charge at the interfaces. 
 The lattice screening provided by KNbO$_{3}$ is much stronger than the 
 one anticipated in LaAlO$_{3}$, where the structural distortion 
 sustains the insulating behavior up to only 5 overlayers of LaAlO$_{3}$ 
 on SrTiO$_{3}$.~\cite{Pentcheva-09}
 As a consequence, in our simulations all the computed structures 
 are insulating.
 The reason behind this is that the LaAlO$_{3}$ is a wide band gap insulator 
 with a low dielectric constant ($\epsilon_{r} = 25$)
 and no ferroelectric instability ($P^{\rm S}_{\rm LaAlO_{3}}$ =0).
 It costs some energy to polarize it.
 On the contrary, KNbO$_{3}$ is a ferroelectric oxide that   
 polarizes spontaneously and contributes to reduce the field 
 [making smaller the numerator of Eq.~(\ref{eq:fieldpol})],
 and increase the critical thickness for the formation of the 2DEG
 [making larger the denominator of Eq.~(\ref{eq:crithick})].
 The reduction of the internal field within KNbO$_{3}$ can be directly 
 checked from the nanosmoothed~\cite{Colombo-91,Junquera-07} 
 electrostatic potential
 in the slabs. Independently of the geometry,
 the magnitude of the field amounts to 0.024 eV/{\AA} [see red dashed lines in 
 Fig.~\ref{fig:potential}]
 to be compared with the roughly constant electric field of 0.24 eV/{\AA}
 (one order of magnitude larger) found in LaAlO$_{3}$/SrTiO$_{3}$ 
 interfaces.~\cite{Lee-08,Cancellieri-11}
 This is in very good agreement with the prediction of the electrostatic model
 developed in Sec.~\ref{sec:summaryer}: using a dielectric constant
 of bulk KNbO$_{3}$ around the ferroelectric structure
 under the same mechanical boundary condition of 
 25.0,~\cite{Stengel-private} 
 Eq.~(\ref{eq:fieldpol}) yields a value of 0.023 eV/{\AA}.

 Since the interfacial potential steps between KNbO$_{3}$ and BaTiO$_{3}$ 
 are 1.08 eV for the TiO$_{2}$/KO interface
 and 1.31 eV for the BaO/NbO$_{2}$ interface 
 (both values computed using the recipe given in Ref.~\onlinecite{Stengel-09.2}
 when the bands of one material are tilted), 
 the estimated critical thickness to trigger the polar catastrophe 
 under the condition of a vanishing electric displacement 
 according to Eq.~(\ref{eq:crithick})
 are, respectively, 170 \AA\ and 180 \AA\
 (between 42 and 44 unit cells), one order of magnitude
 larger than in LaAlO$_{3}$/SrTiO$_{3}$.
  
\subsection{In-plane polarization} 
\label{sec:in-plane}

 Even though the ground state of bulk KNbO$_{3}$ and BaTiO$_{3}$ 
 at zero temperature is rhombohedral, 
 where the polarization displays both in-plane $(x,y)$ and 
 out-of-plane ($z$) components, 
 when a compressive in-plane strain is applied
 the polarization in the $(x,y)$-directions is strongly reduced
 or even suppressed.~\cite{Dieguez-04} 
 In particular, when KNbO$_{3}$ or BaTiO$_{3}$ thin films are grown on top of
 a SrTiO$_{3}$ substrate, the in-plane polarization of these materials 
 vanishes and the tetragonal $c$-phase is
 stabilized.~\cite{Yoneda-98.1,Yoneda-98.2} 
 However, contrary to current thought, in our calculations 
 of the BaTiO$_{3}$/KNbO$_{3}$ interfaces, 
 we can observe in Fig.~\ref{fig:rumpling} the appearance of a moderate 
 in-plane polarization in these systems along the [110] direction. 
 The layer-by-layer in-plane rumpling profile
 plotted in Fig. \ref{fig:rumpling} 
 reveals that the effect is highly localized at the 
 $p$-type TiO$_2$/KO interface,
 quickly decaying upon moving into BaTiO$_{3}$, 
 and it is completely absent at the $n$-type BaO/NbO$_2$ interfaces. 

\begin{figure} [h]
    \begin{center}
       \includegraphics[width=6.5cm]{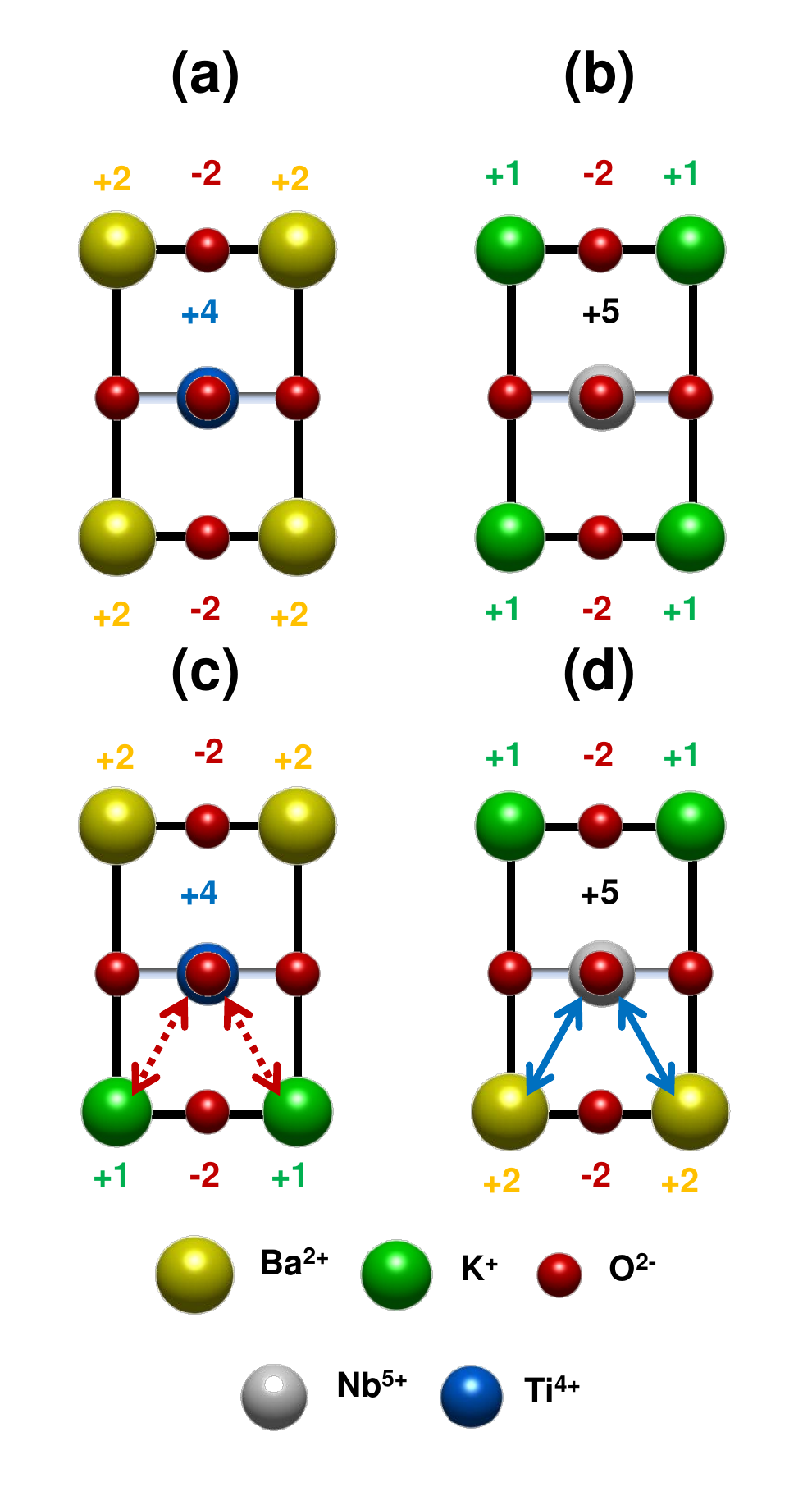}
       \caption{(Color online) Schematic view of bulk unit cells
                of (a) BaTiO$_{3}$ and (b) KNbO$_{3}$,
                together with the ideal atomic structure at
                (c) $p$-type TiO$_{2}$/KO, and
                (d) $n$-type BaO/NbO$_{2}$ interfaces.
                Atoms are represented by balls following the same 
                conventions as in Fig.~\ref{fig:rumpling}.
                Numbers indicate the nominal charge of the different ions.
                Solid blue (dashed red) arrows represent an 
                enhancement (depletion) of the in-plane 
                electrostatic repulsion between ions.
                The schema shows how the replacement of ions 
                favors the movement of the transition metal B-cation in 
                TiO$_2$/KO interfaces and hinders it in the BaO/NbO$_2$ 
                interfaces.}
       \label{fig:inplanepol}
    \end{center}
 \end{figure}

 The origin of the hidden interfacial in-plane polarization 
 can be easily understood with a simple electrostatic model based on 
 formally charged ions.
 In Fig.~\ref{fig:inplanepol}  we compare the atomic structure of 
 bulk BaTiO$_{3}$ [Fig.~\ref{fig:inplanepol}(a)] 
 and KNbO$_{3}$ [Fig.~\ref{fig:inplanepol}(b)] with that present in the 
 $p$-type TiO$_{2}$/KO [Fig.~\ref{fig:inplanepol}(c)] 
 and $n$-type BaO/NbO$_2$ [Fig.~\ref{fig:inplanepol}(d)] interfaces. 
 The cleavage of bulk BaTiO$_{3}$ and KNbO$_{3}$ to form the 
 $p$-type TiO$_{2}$/KO interface is accompanied by a change in the local
 electrostatic potential felt by the Ti cations.
 At the interface, some of the Ba cations (nominal charge +2) in the 
 first neighbour atomic layer are replaced by K cations (nominal charge +1).
 This implies that at that interface the in-plane Ti cation movement 
 is less constrained than in bulk, due to the reduced repulsions 
 with K ions with respect to Ba ones. 
 Analogously, in the $n$-type interface the electrostatic potential
 felt by the Nb atoms is also altered. In this case the K$^{+1}$ cations
 are replaced by Ba$^{+2}$ ions, leading to an enhanced 
 repulsion that hinders the appearance of any in-plane polarization. 
 As this effect is directly related to the nominal charges of the ions in 
 both interfacial materials we would expect it to be present in other 
 similar systems and could be used to induce in-plane polarizations in
 other ferroelectric nanostructures.

\begin{figure} [h]
    \begin{center}
       \includegraphics[width=8.5cm]{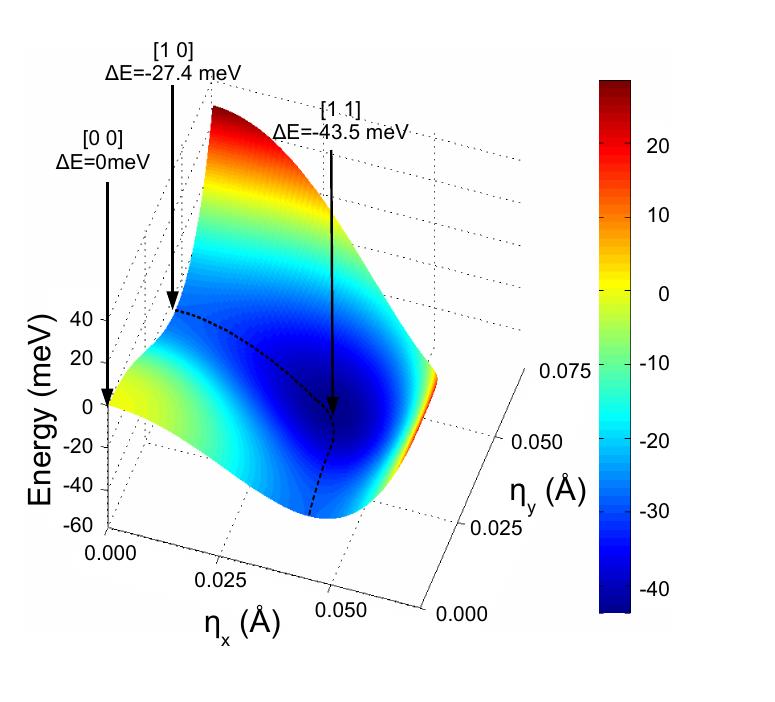}
       \caption{(Color online) Two-dimensional energy surface as a
                function of the in-plane distortions of the
                BaTiO$_{3}$ atoms, as indicated in the main text.
                The dashed line follows the minimum of the valley.
                Units in meV.}
        \label{fig:ensurf}
    \end{center}
 \end{figure}

 To further characterize the development of the in-plane polarization,
 we take the relaxed coordinates of the asymmetric slab with
 $m$ = 5 and $l$ = 8 and remove by hand the in-plane displacements.
 We use this new structure as a reference configuration.
 Then, we compute the in-plane distortions required to go
 from the reference configuration to the relaxed structure and
 decompose into their $x$ and $y$ components.
 Finally, given fractions of these distortions
 are frozen in on top of the reference structure.
 The energy surface obtained for different fractions of the in-plane
 rumplings is represented in Fig.~\ref{fig:ensurf}. 
 From this energy landscape we can extract (i) the stabilization energy 
 (the energy difference between
 the polar state and the reference one), that amounts 
 to 43.5 meV/slab, and 
 (ii) the energy barriers that prevents the system to rotate 
 the in-plane polarization from [110] direction to any of the 
 other symmetry-equivalent positions, passing through the transition state at 
 [100] positions. This is only 16 meV, slightly larger than those found in  
 artificial Ruddlesden-Popper-type superlattices.~\cite{Nakhmanson-10} 
 To gauge the magnitude of these barriers, we compare them with 
 the rotation zero-point-energy (ZPE) in 
 this slab. This value is obtained through the nuclear motion 
 Hamiltonian

 \begin{equation}
    \hat{H} = -\frac{\hbar^2}{2M^*}\left(\frac{\partial^2}{\partial \eta_x^2}
       +\frac{\partial^2}{\partial \eta_y^2}\right)+V\left(\eta_x,\eta_y\right),
    \label{eq:nucmotion}
 \end{equation}

 \noindent where $V\left(\eta_x,\eta_y\right)$ is the energy 
 represented in Fig.~\ref{fig:ensurf}
 and $M^*$ can be shown to be

 \begin{equation}
   \frac{1}{M^*}=\sum_{i}\frac{c_i^2}{M_i},
   \label{eq:mass}
 \end{equation}

 \noindent by writing the kinetic energy operator expressed 
 in terms of atomic coordinates and
 masses ($M_i$) using the linear trasformation relating the effective modes
 $\eta_x$ and $\eta_y$ with the atomic displacements through the 
 coefficients $c_i$.
 To solve the Schr\"odinger equation associated to Eq.~(\ref{eq:nucmotion}) and 
 find the vibrational 
 levels associated to Fig.~\ref{fig:ensurf} we follow 
 the recipe given in Ref.~\onlinecite{Garcia-Fernandez-10}. Then,
 we compute the ZPE by comparing the position of the first level 
 to the minimum of the energy surface.
 The resulting value for the ZPE, 13.6 meV, is much smaller than the global
 stabilization energy but comparable to the height of the rotational barrier,
 indicating that 
 quantum fluctuations will be important in these nanostructures.
 These fluctuations prevent the observation of any 
 in-plane spontaneous polarization,  
 since the system will be delocalized over the four equivalent minima 
 in [110] directions in a similar way to what happens in a dynamic Jahn-Teller 
 problem.~\cite{Garcia-Fernandez-10} 

 This case can be compared with what happens in an incipient ferroelectric. In 
 the latter case quantum fluctuations make the ZPE larger than the double well
 stabilization energy making the maximum of the probability 
 density associated to the distortion (and the polarization) to be 
 localized at the centrosymmetric state  
 (origin in Fig.~\ref{fig:ensurf}). 
 On the other hand, in the present case, the 
 maximum of the probability density corresponds with a non-zero polarization
 region around the origin (see dashed line in Fig.~\ref{fig:ensurf}) 
 but directional averaging results in a null
 net polarization.
 However, the coherent dynamics between the wells could be disrupted  
 by small electric fields along the plane, that would induce large  
 changes in the directionality of the in-plane polarization. 
 The signature of such potential energy surface for polarization
 rotation would be a high dielectric constant.~\cite{Nakhmanson-10}

\section{Conclusions}
\label{sec:conclusions}

 In summary,
 using accurate first-principles simulations we have
 studied the influence of the ferroelectric polarization of the 
 polar layer in the formation of 2DEG at 
 BaTiO$_{3}$/KNbO$_{3}$ interfaces.
 The most important conclusions that can be drawn are:
 (i) the spontaneous polarization of the polar KNbO$_{3}$ layer cancels
 out almost exactly the ``built-in'' polarization discontinuity at the
 interface;
 (ii) as a consequence of this compensation, the critical thickness
 for the formation of the 2DEG is estimated to be between 42 and 44 unit cells
 of KNbO$_{3}$, one order of magnitude larger than in 
 SrTiO$_{3}$/LaAlO$_{3}$ interfaces;
 (iii) this behavior can be easily explained in terms of a simple
 model based on the modern theory of polarization and basic electrostatics;
 and (iv) surprisingly, BaTiO$_{3}$ displays an in-plane component
 of the polarization at the $p$-type TiO$_{2}$/KO interface,
 even when BaTiO$_{3}$ is under in-plane compressive strains.  
 However we do not expect this in-plane polarization to be experimentally
 observable as the barriers for its rotation are very small and quantum 
 fluctuations will prevent it from being localized in a particular direction. 
 This situation could be easily modified by the application of small electric 
 fields that will break the symmetry and reveal the hidden polarization. 


 We thank Massimiliano Stengel for useful discussions and a critical
 reading of the manuscript.
 This work was supported by the Spanish Ministery of Science and
 Innovation through the MICINN Grant FIS2009-12721-C04-02, by the
 Spanish Ministry of Education through the FPU fellowship AP2006-02958 (PAP),
 and by the European Union through the project EC-FP7,
 Grant No. CP-FP 228989-2 ``OxIDes''.
 The authors thankfully acknowledge the computer resources,
 technical expertise and assistance provided by the
 Red Espa\~nola de Supercomputaci\'on.
 Calculations were also performed at the ATC group
 of the University of Cantabria.


\bibliographystyle{apsrev4-1}

\begin{thebibliography}{60}
\bibitem [{Del()}]{Deltap}%
  \BibitemOpen
  \href@noop {} {}\bibinfo {note} {For a $p$ interface of type II, $\Delta =
  E_{\rm gap}^{\rm non-polar} - \left( E_{\rm CBM}^{\rm non-polar} - E_{\rm
  CBM}^{\rm polar}\right)$, where $E_{\rm CBM}$ is the conduction band minimum
  of the corresponding layer right at the interface.}\BibitemShut {Stop}%
\end{thebibliography}

\end{document}